\documentclass[11pt,a4paper,english]{amsart}
\usepackage{babel}
\usepackage[latin1]{inputenc}
\usepackage{amsmath}
\usepackage{amsthm}
\usepackage{amsfonts}
\usepackage{indentfirst}
\usepackage{graphicx}
\usepackage{amssymb}
\usepackage[mathscr]{eucal}
\usepackage{tikz}
\usepackage{caption}
\usepackage{enumerate}
\usepackage{amsthm}
\usepackage{amscd}
\newtheorem{teo}{Theorem}[section]

\theoremstyle{definition}
\newtheorem{rem}[teo]{Remark}

\title[On the generalization of the construction of quantum codes]{On the generalization of the construction of quantum codes from Hermitian self-orthogonal codes}
\author{Carlos Galindo and Fernando Hernando}
\curraddr{Instituto
Universitario de Matem\'aticas y Aplicaciones de Castell\'on and
Departamento de Matem\'aticas, Universitat Jaume I, Campus de Riu
Sec. 12071 Castell\'{o} (Spain).
}
\email{{\rm Galindo:} galindo@uji.es; {\rm Hernando:} carrillf@uji.es}
\date{}
\thanks{Partially supported by the Spanish Government MICINN/FEDER/AEI/UE, grants  PGC2018-096446-B-C22, RED2018-102583-T, as well as by Universitat Jaume I, grant UJI-2018-10.}


\begin{document}

\begin{abstract}
 Many $q$-ary stabilizer quantum codes can be constructed from Hermitian self-orthogonal $q^2$-ary linear codes. This result can be generalized to $q^{2 m}$-ary linear codes, $m > 1$. We give a result for easily obtaining quantum codes from that generalization. As a consequence we provide several new binary stabilizer quantum codes which are records according to \cite{codet} and new $q$-ary ones, with $q \neq 2$, improving others in the literature.
\end{abstract}

\maketitle

\section*{Introduction}
The importance of quantum information processing is beyond doubt due to its spectacular applications.
Nowadays there is some evidence of quantum processors capable of executing certain tasks greatly improving classical processors \cite{Aru} which increases the interest in  tools for the proper functioning of quantum computers such as the quantum error-correcting codes (QECCs). QECCs are mainly designed for protecting quantum information from quantum noise and decoherence. Notice that, despite quantum information cannot be cloned \cite{8AS, 26RBC}, quantum error correction works \cite{23RBC, 95kkk}. These facts explain why many researchers are interested in obtaining QECCs with good parameters (which measure the behaviour of the codes) and the literature contains a large quantity of papers devoted to finding QECCs with better parameters than others previously obtained.

Let $q$ be a  prime power, a $q$-ary QECC of length $n$ is a subspace of the Hilbert space $\mathcal{H}=\mathbb{C}^{q^n}$. The most used class of quantum codes are stabilizer quantum codes. They are obtained as the intersection of the eigenspaces, corresponding to the eigenvalue $1$, of the elements of some subgroup of the error group generated by a suitable error basis of the Hilbert space $\mathcal{H}$. The parameters of a QECC,  length, dimension and minimum distance, are usually denoted by $((n,K,d))_q$, where errors with weight less than $d$ either can be detected or have no effect on $C$ but some error with weight $d$ cannot be detected. We are only interested in $q^k$-dimensional subspaces of $\mathcal{H}$ and, abusing of notation and when no confusion arises, we say that these QECCs have dimension $k$; in this case, the parameters are usually written as $[[n,k,d]]_q$.

QECCs were firstly introduced in the binary case, where one finds the seminal papers on the subject \cite{20kkk, 38kkk, 18kkk, 19kkk, 45kkk, 7kkk, 8kkk}. Later QECCs were studied for the general $q$-ary case (see \cite{BE, 71kkk, AK,  35kkk, opt, kkk, Akk, lag3, lag1, jin, lag2, SLLG, galindo-hernando, gal-her-rua, QINP2, traza, cao-cui2, cao-cui} among many other articles). The general case is particularly interesting for fault-tolerant computing \cite{FTShor, FTKnill, FTPres, FTGot, FTSte, FTCam, luol}.

One of the main advantages of stabilizer codes is that their existence is equivalent to that of self-orthogonal additive codes with respect to certain trace-symplectic form (see \cite{AK} or \cite[Theorem 13]{kkk}). This trace-symplectic form is not very used but the above result allows us to deduce  that many stabilizer quantum codes can be derived from self-orthogonal classical codes with respect to the Hermitian or the Euclidean inner product. Usually one finds good stabilizer codes over $\mathbb{F}_q$ by considering Hermitian self-orthogonal codes over $\mathbb{F}_{q^2}$. The specific result, Theorem \ref{La1},  shows that an $[[n,n-2k, \geq d^{\perp_h}]]_q$ quantum code can be constructed from a Hermitian self-orthogonal $[n,k]_{q^2}$ linear code $C$ over $\mathbb{F}_{q^2}$, where $d^{\perp_h}$ stands for the minimum distance of the Hermitian dual code $C^{\perp_h}$. This result has been extensively used in many papers to give many good QECCs \cite{lag3, lag2, jin, zh}.

In the present paper we recall that Theorem \ref{La1} can be regarded as a special case of a more general result by considering linear codes over certain extensions of $\mathbb{F}_{q}$. Indeed, Theorem \ref{Elbueno} states that if $C$ is a linear code over $\mathbb{F}_{q^{2 m}}$, $m \geq 1$, with parameters $[n,k]_{q^{2 m}}$ which is self-orthogonal with respect to the Hermitian inner product, then there exists an $[[m n, m n - 2 m k, \geq d^{\perp_h}]]_q$ stabilizer quantum code. Theorem \ref{Elbueno} is a straightforward consequence of \cite[Lemma 76]{kkk} which seems to have gone unnoticed by many researchers because, in the literature, we have not found new quantum codes considering $m>1$.

We expect that many good quantum codes can be established by this result. Our goal is to give some evidence by stating (and proving) Theorem \ref{elH} which combined with Theorem \ref{Elbueno} gives rise to a number of stabilizer quantum codes with good parameters. Theorem \ref{elH} derives from \cite{jin} and gives an easy way to find Hermitian self-orthogonal codes. The above mentioned combination produces, in the binary case, new QECCs which are records according to \cite{codet}. Here, the word record means we provide codes for entries in \cite{codet} whose constructions were missing. There is no collection of tables as \cite{codet} for non-binary QECCs but one can find many papers in the literature about them. Most of these papers are devoted to quantum MDS codes which have relatively small length \cite{chen, refer, zh}. Since we are able to construct long QECCs, we use recent articles \cite{SLLG, cao-cui, LV} for comparison and show that with our method we can improve the parameters of a number of codes therein.

Section \ref{secuno} of the paper is devoted to recall  Theorem \ref{Elbueno} for obtaining QECCs from linear codes. Theorem \ref{elH} and parameters (some of them displayed in tables) of new QECCs can be found in Section \ref{secdos}. As mentioned all the provided parameters correspond to QECCs obtained by applying Theorems \ref{elH} and \ref{Elbueno}. In the binary case, our results together with propagation rules determine $91$ new QECCs which are records according to \cite{codet}. We use the rules that state that the existence of an $[[n,k, d]]_q$ quantum code implies that of an $[[n+1,k, \geq d]]_q$ quantum code (lengthening) and,  also, that of an $[[n,k-1, d]]_q$ quantum code (subcode-construction) \cite[Lemmas 69 and 71]{kkk}.

\section{A construction of stabilizer quantum codes}
\label{secuno}
Let $q=p^r$, where $p$ is a prime and $r$ a positive integer. Many good stabilizer quantum codes over the finite field with $q$ elements $\mathbb{F}_q$ are obtained from linear codes over the finite field $\mathbb{F}_{q^2}$ which are self-orthogonal under the Hermitian inner-product. Recall that given two vectors $\boldsymbol{x} = (x_1, x_2, \ldots, x_n)$ and $\boldsymbol{y} = (y_1, y_2, \ldots, y_n)$ in $\mathbb{F}_{q^2}^n$, $n \geq 1$, their {\it Hermitian inner product} is defined by
\[
\boldsymbol{x} \cdot_h \boldsymbol{y} := \sum_{i=1}^n x_i y_i^q \in \mathbb{F}_{q^2},
\]
and the specific result to construct stabilizer quantum codes is the following (see \cite[Corollary 16 and Lemma 18]{kkk}).
\begin{teo}
\label{La1}
Let $C$ be  an $\mathbb{F}_{q^2}$-linear code of length $n$ and dimension $k$. Assume that $C$ is Hermitian self-orthogonal, i.e.
\[
C \subseteq C^{\perp_h} := \left\{ \boldsymbol{x} \in \mathbb{F}_{q^2}^n \; | \; \boldsymbol{x} \cdot_h \boldsymbol{y} =0 \mbox{ for all $\boldsymbol{y}$ in $C$} \right\}.
\]
Then, there exists a stabilizer quantum code over $\mathbb{F}_{q}$ with parameters $[[n,n-2k, \geq d^{\perp_h}]]_q$, where $d^{\perp_h}$ stands for the minimum distance of the code $C^{\perp_h}$.
\end{teo}

Next we recall a generalization of Theorem \ref{La1} allowing the use of codes over extension fields of $\mathbb{F}_{q^2}$. We will prove that one can obtain long stabilizer codes with good parameters over $\mathbb{F}_{q}$ by considering linear codes over fields $\mathbb{F}_{q^{2 m}}$, $m >0$, which are self-orthogonal with respect to the Hermitian inner product.



\begin{teo}
\label{Elbueno}
Let $C$ be an $\mathbb{F}_{q^{2m}}$-linear code of length $n$ and dimension $k$, $m>0$. Suppose that $C \subseteq C^{\perp_h}$, where
\[
C^{\perp_h} := \left\{  \boldsymbol{x} \in \left(\mathbb{F}_{q^{2 m}}\right)^n \; | \; \boldsymbol{x} \cdot_h \boldsymbol{y} := \sum_{i=1}^n x_i y_i^{q^{m}} = 0 \mbox{ for all $\boldsymbol{y}$ in $C$}
\right\}.
\]
Then, there exists an $\mathbb{F}_{q}$-stabilizer quantum code with parameters
\[
\big[\big[ m n, mn - 2 m k, \geq d^\perp_h  \big]\big]_q,
\]
where $d^\perp_h$ is the minimum distance of the code $C^{\perp_h}$.
\end{teo}

This result can be deduced from Lemma 76 in \cite{kkk} (see also \cite{AK}) which states that the existence of an $((n,K,d))_{q^m}$ stabilizer code implies that of an $((mn,K, \geq d))_q$ stabilizer code. Then, to deduce Theorem \ref{Elbueno}, it suffices to consider the above code $C$ to obtain an $[[n, n- 2k, \geq d^{\perp_h}]]_{q^m}$ stabilizer code by applying  Theorem \ref{La1} and by \cite[Lemma 76]{kkk} there exists a $q$-ary stabilizer code as in the statement.

Surprisingly we have not found in the literature new quantum codes obtained from Theorem \ref{Elbueno}, $m>1$. We think that this result goes unnoticed by many researchers. We learned it when a reviewer pointed out to us the existence of \cite[Lemma 76]{kkk} after a first version of this paper where, in a different way,  we proved Theorem \ref{Elbueno} for the particular case where $m=2^{\ell -1}$, $\ell \geq 1$.

\section{Hermitian self-orthogonal codes and examples}
\label{secdos}
We devote this section to show that some very good stabilizer quantum codes can be derived  from Theorem \ref{Elbueno}.


\subsection{A useful result}

Next we state and prove a result derived from \cite[Theorem 2.5]{jin} which provides suitable linear codes to apply Theorem \ref{Elbueno}. This procedure gives some binary stabilizer quantum codes which are records according to \cite{codet} (in the sense explained in the introduction) and also some $q$-ary stabilizer quantum codes, $q \neq 2$, improving the parameters of codes in the recent literature.

We start by recalling Theorem 2.5 in \cite{jin}.
\begin{teo}
\label{elF}
Let $e$ be a prime power and set $Q=e^2$. Consider an integer $2 \leq n \leq Q$ and write
$n=n_1 + n_2 + \cdots + n_t$, where $1 \leq t \leq e$ and $2 \leq n_i \leq e$ for all $1 \leq i \leq t$. Then, for any positive integer
\[
1 \leq k  \leq \frac{\min\{n_1, n_2, \ldots, n_t\}}{2},
\]
there exists an $[n,k]_Q$ linear code $\mathcal{C}$ over the field $\mathbb{F}_Q$ which is Hermitian self-orthogonal and the minimum distance of $\mathcal{C}^{\perp_h}$ is $k+1$.
\end{teo}

Next we state a new result which will be useful.

\begin{teo}
\label{elH}
Let $e>2$ be a prime power and set $Q:= e^2$. Consider an integer $2 \leq n \leq Q$ and write $n$ as $n=a e + b$, where $0 \leq a < e$ and $0 \leq b < e$, i.e. the $e$-adic expression of $n$, and include the case $a=e$ and $b=0$.

Define $K_n$ as follows: $K_n:= \lfloor e/2 \rfloor$ when $b=0$. $K_n:= \lfloor n/2 \rfloor$ when $a=0$. $K_n:= \lfloor (e-1) /2 \rfloor $ when $a \neq 0$, $b \neq 0$ and $a + b \geq e$. Otherwise,
\[
K_n := \left\lfloor \frac{\max \left\{ \lfloor n/(a+1) \rfloor, a +b   \right\} }{2} \right\rfloor.
\]
Then, for each $1 \leq k \leq K_n$, there exists an $[n,k]_Q$ linear code $\mathcal{C}$ which is self-orthogonal for the Hermitian inner product and such that the minimum distance of the Hermitian dual code $\mathcal{C}^{\perp_h}$ is $k+1$.
\end{teo}

\begin{proof}
Assume $b=0$, then the result holds by setting $n_1=n_2 = \cdots = n_a = e$ and applying Theorem \ref{elF}. When $a=0$, the same theorem with $n_1=n$ proves the result.

Suppose $a \neq 0 \neq b$ and $a + b \geq e$. Let us see that there exist non-negative integers $i, j$ such that $i+j=a$ and positive integers $n_1 = \cdots = n_i = e$, $n_{i+1} = \cdots =n_{i+j}=e -1$ and $n_{a+1}= e -1$ that are suitable to apply Theorem \ref{elF}, which concludes the result in this case. Indeed,
\[
ie + j(e -1) + e -1 = (i+j) e + e -1 -j = a e + e -1 -j =n,
\]
for some $j$ because the fact that $a+b \geq e$ proves the existence of such a $j$ with $0 \leq j \leq e-1$.

Finally, assume $a +b < e$. Then, on the one hand, setting $n_1=n_2 = \cdots = n_a = e -1$ and $n_{a+1} = a +b$, we find the second bound for $k$, $\lfloor (a+b)/2 \rfloor$, by Theorem \ref{elF}. With respect to the first one, it is clear that
\[
(a+1) \left\lfloor  \frac{n}{a+1} \right\rfloor \leq n \leq (a+1) \left(  \left\lfloor  \frac{n}{a+1} \right\rfloor +1 \right).
\]
This implies that a set $\{n_i\}_{i=1}^{a+1}$ as in Theorem \ref{elF} can be constructed for values $n_i$ which are either $\left\lfloor  \frac{n}{a+1} \right\rfloor$ or $\left\lfloor  \frac{n}{a+1} \right\rfloor + 1$. This concludes the proof because we can choose the best bound.
\end{proof}

\subsection{Examples}
In  this subsection we determine parameters of some (constructible) stabilizer quantum codes over finite fields of small cardinality.

\subsubsection{Binary stabilizer quantum codes} We start with binary codes. In this case we give a number of quantum codes which are records according to \cite{codet}, that is we give codes for entries in \cite{codet} whose constructions were missing. We explain in detail how we get our first binary code.

We start with the values $e = 16$ and $Q=16^2 = 256$. By Theorem \ref{elH}, we can consider the value $n=63$ because $2 \leq 63 \leq 256$ and $63=n= 3 \cdot 16 + 15$. Thus $a=3$ and $b=15$. Since $a+b \geq 16$, $K_{63}= \lfloor 15 / 2 \rfloor =7$. Then, by Theorem \ref{elH}, there exists a suitable linear code $\mathcal{C}$ over $\mathbb{F}_Q$ with length $n=63$, dimension $k=6$ and $d(\mathcal{C}^{\perp_h}) =7$. Now $Q=q^{2m}$ for $q=2$ and $m=4$. Applying Theorem \ref{Elbueno}, we get a
$\mathbf{[[252,204, \geq 7]]_2}$ quantum code which is a record. If we pick $n=62$ and $k=7$, then a new record is obtained: $\mathbf{[[248,192, \geq 8]]_2}$.

With an analogous procedure we obtain new records according to \cite{codet}. By using either lengthening or subcode-construction (the two propagation rules described at the end of the introduction) we get  more records. All of them are grouped in Table \ref{tabla1}, where the parameters obtained without propagation rules are marked with a * and those obtained by lengthening (respectively, subcode-construction) are marked with an $L$ (respectively, $S$).
\begin{table}
\begin{center}
\begin{tabular}{||c|c|c||c|c|c||c|c|c||c|c|c||}
\hline
  $n$ & $k$ & $ \geq d$ & $n$ & $k$ & $\geq d$ & $n$ & $k$ & $\geq d$  & $n$ & $k$ & $\geq d$\\
  \hline
  252* & 204 & 7 & 252* & 196 & 8 & 248* & 200 & 7 & 248* & 192 & 8\\
\hline
244* & 196 & 7 & 244* & 188 & 8 & 240* & 192 & 7 & 240* & 184 & 8\\
\hline
$252^S$  & 203 & 7 & $252^S$ & 195 & 8 & $251^L$ & 200 & 7 & $251^S$ & 199 & 7\\
\hline
$251^S$  & 198 & 7 & $251^L$ & 192 & 8 & $251^S$ & 191 & 8 & $251^S$ & 190 & 8\\
\hline
$250^L$ & 200 & 7 & $250^S$ & 199 & 7 & $250^S$ & 198 & 7 & $250^S$ & 197 & 7\\
\hline
$250^L$ & 196 & 7 & $250^L$ & 192 & 8 & $250^S$ & 191 & 8 & $250^S$ & 190 & 8\\
\hline
$249^L$ & 200 & 7 & $249^S$ & 199 & 7 & $249^S$ & 198 & 7 & $249^S$ & 197 & 7\\
\hline
$249^S$ & 196 & 7 & $249^S$ & 195 & 7 & $249^L$ & 192 & 8 & $249^S$ & 191 & 8\\
\hline
$249^S$ & 190 & 8 & $249^S$ & 189 & 8 & $248^S$ & 199 & 7 & $248^S$ & 198 & 7\\
\hline
$248^S$ & 197 & 7 & $248^S$ & 196 & 7 & $248^S$ & 195 & 7 & $248^S$ & 194 & 7\\
\hline
$248^S$ & 191 & 8 & $248^S$ & 190 & 8 & $248^S$ & 189 & 8 & $248^S$ & 188 & 8\\
\hline
$247^L$ & 196 & 7 & $247^S$ & 195 & 7 & $247^S$ & 194 & 7 & $247^S$ & 193 & 7\\
\hline
$247^L$ & 188 & 8 & $247^S$ & 187 & 8 & $246^L$ & 196 & 7 & $246^S$ & 195 & 7\\
\hline
$246^S$ & 194 & 7 & $246^S$ & 193 & 7 & $246^S$ & 192 & 7 & $246^L$ & 188 & 8\\
\hline
$246^S$ & 187 & 8 & $246^S$ & 186 & 8 & $245^L$ & 196 & 7 & $245^S$ & 195 & 7\\
\hline
$245^S$ & 194 & 7 & $245^S$ & 193 & 7 & $245^S$ & 192 & 7 & $245^S$ & 191 & 7\\
\hline
$245^L$ & 188 & 8 & $245^S$ & 187 & 8 & $245^S$ & 186 & 8 & $245^S$ & 185 & 8\\
\hline
$244^S$ & 195 & 7 & $244^S$ & 194 & 7 & $244^S$ & 193 & 7 & $244^S$ & 192 & 7\\
\hline
$244^S$ & 191 & 7 & $244^S$ & 187 & 8 & $244^S$ & 186 & 8 & $244^S$ & 185 & 8\\
\hline
$244^S$ & 184 & 8 & $243^L$ & 192 & 7 & $243^S$ & 191 & 7 & $243^L$ & 184 & 8\\
\hline
$243^S$ & 183 & 8 & $242^L$ & 192 & 7 & $242^S$ & 191 & 7 & $242^L$ & 184 & 8\\
\hline
$242^S$ & 183 & 8 & $241^L$ & 192 & 7 & $241^S$ & 191 & 7 & $241^L$ & 184 & 8\\
\hline
$241^S$ & 183 & 8 & $240^S$ & 191 & 7 & $240^S$ & 183 & 8 & - & - & -\\
\hline

\end{tabular}
\end{center}
\caption{Binary stabilizer quantum records}
\label{tabla1}
\end{table}


\subsubsection{Non-binary stabilizer quantum codes}
As we did in the binary case, we explain in detail the construction of some families of good stabilizer quantum codes over $\mathbb{F}_4$. We will only show the parameters of the remaining stabilizer codes which can be obtained in a similar way.

Set $e=16$, $Q=16^2 = 256$ and, applying Theorem \ref{elH}, pick $n=76$, which accomplishes $2 \leq 76 \leq 256$. Now $n=76=4 \cdot 16 + 12$, then under the notation of that theorem $a=4$ and $b=12$. Since $a + b \geq 16$, $K_{76} = 7$, and there exists a linear code over $\mathbb{F}_Q$ which is self-orthogonal for the Hermitian inner product, where $Q=q^{2m}$ for $q=4$ and $m=2$. Applying Theorem \ref{Elbueno}  we find the stabilizer quantum codes over $\mathbb{F}_4$ showed in Table \ref{tabla2}.

\begin{table}
\begin{center}
\begin{tabular}{||c|c|c||c|c|c||}
\hline
$n$ & $k$ & $\geq d$ & $n$ & $k$ & $\geq d$ \\
\hline
152 &148 &2&152 &144 &3 \\ \hline
152 &140 &4&152 &136 &5 \\ \hline
152 &132 &6&152 &128 &7 \\ \hline
152 &124 &8&- &- &- \\
 \hline
\end{tabular}
\end{center}
\caption{Stabilizer quantum codes over $\mathbb{F}_4$ }
\label{tabla2}
\end{table}

By lengthening, we obtain quantum codes with parameters $$[[153,140,\geq 4]]_4,[[153,136,\geq 5]]_4, \;\; [[153,132, \geq 6]]_4 \;\; [[153,128,\geq 7]]_4$$
improving some codes in (and adding a new one to) \cite[Table 3]{SLLG}.

Looking for more $4$-ary stabilizer codes, set $q=4$ and $m=3$. Write $Q=q^{2m}$ and $e=q^m=64$. Pick $n=255$ which gives $a=3$ and $b=63$ with the notation of Theorem \ref{elH}. Then $K_{255}=31$ and applying Theorems \ref{elH} and \ref{Elbueno} one gets a family of stabilizer codes with parameters
\[
\left\{ [[765, 765 - 6j, \geq 1+j]]_4 \right\}_{18 \leq j \leq 31}.
\]
These codes improve a lot some given in \cite[Table 2]{SLLG} whose minimum distance $d$ satisfies $19 \leq d \leq 32$. For instance we give codes with parameters $[[765, 657, \geq 19]]_4$, $[[765, 651, \geq 20]]_4$ and $[[765, 645, \geq 21]]_4$ while the parameters of the corresponding codes in \cite{SLLG} are $[[765, 643, \geq 19]]_4$, $[[765, 639, \geq 20]]_4$ and $[[765, 631, \geq 21]]_4$.

We conclude this section by giving  some more families of stabilizer quantum codes obtained with our procedure.

We start with a $3$-ary stablizer quantum code with parameters $[[110,98, \geq 4]]_3$ which improves the quantum code with parameters $[[110,96, \geq 4]]_3$ given in \cite{LV}. Our code is obtained by setting, with the previous notation, $n=55$, $q=3$ and $m=2$.

Similarly, considering $n=234$, we get stabilizer quantum codes over $\mathbb{F}_5$ with parameters as in Table \ref{tabla3}.
\begin{table}
\begin{center}
\begin{tabular}{||c|c|c||c|c|c||}
\hline
$n$ & $k$ & $\geq d$ & $n$ & $k$ & $\geq d$ \\ \hline
468 &452 &5&468 &448 &6 \\ \hline
468 &444 &7&468 &440 &8 \\ \hline
468 &436 &9&468 &432 &10 \\  \hline
468 &428 &11&468 &424 &12 \\
 \hline
\end{tabular}
\end{center}
\caption{Stabilizer quantum codes over $\mathbb{F}_5$ }
\label{tabla3}
\end{table}
Notice that these codes make a great improvement to some ones in \cite[Table 4]{SLLG}.

Now we provide the parameters of a family of QECCs over $\mathbb{F}_7$. Consider $Q=2401 = 7^4$ and $n=196$, again by Theorems \ref{elH} and \ref{Elbueno} we get a family of stabilizer quantum codes with parameters
\[
 \Big\{[[392, 388 - 4j, \geq 2 +j]]_7 \Big\}_{j=3}^{23}.
\]
Comparing with \cite[Table 3]{cao-cui}, we obtain many more $7$-ary quantum codes of length $392$. For $j=3, 4$ our parameters coincide with those in \cite{cao-cui} and we get a $[[392,368, \geq 7]]_7$ code which improves the  $[[392,364, \geq 7]]_7$ code in \cite{cao-cui}.

Our next family corresponds to the field $\mathbb{F}_8$. Theorems \ref{elH} and \ref{Elbueno} for $Q=8^4= 4096$ and $n=283$ give rise to a new family of QECCs with parameters
\[
 \Big\{[[566, 562 - 4j, \geq 2 +j]]_8 \Big\}_{j=5}^{27}.
\]
After lengthening, one gets a set of QECCs with parameters
\[
 \Big\{[[567, 562 - 4j, \geq 2 +j]]_8 \Big\}_{j=5}^{27}.
\]
As before, we add many new codes to those $8$-ary ones in \cite[Table 1]{cao-cui} of length $567$ and obtain a code with parameters $[[567,542, \geq 7]]_8$  improving the $[[567,539, \geq 7]]_8$ code in \cite{cao-cui}.

To end, set $Q=6561=9^4$. As above
\begin{itemize}
\item Picking $n=200$, we get a family of stabilizer quantum codes over $\mathbb{F}_9$ with parameters:
$$ \Big\{[[400, 396 - 4j, \geq 2 +j]]_9 \Big\}_{j=3}^{32}.$$
    \item Setting $n=400$, we obtain: $$ \Big\{[[800, 796 - 4j, \geq 2 +j]]_9 \Big\}_{j=3}^{39}.$$
        \item With $n=405$, we obtain: $$ \Big\{[[810, 806 - 4j, \geq 2 +j]]_9 \Big\}_{j=3}^{39}.$$
            \item Finally, with $n=162$, we get: $$ \Big\{[[324, 320 - 4j, \geq 2 +j]]_9\Big\}_{10 \neq j=7}^{39}.$$
\end{itemize}
With respect to Tables 1, 3, 5 and 8 in \cite{cao-cui} we add quite a few new codes.  In addition we obtain several codes with better parameters than those given in \cite{cao-cui}: $[[400,376, \geq 7]]_9$, $[[800,776, \geq 7]]_9$, $[[800,772, \geq 8]]_9$, $[[810,786, \geq 7]]_9$, $[[810,782, \geq 8]]_9$, $[[810,778, \geq 9]]_9$, $[[324,276, \geq 13]]_9$ and $[[324,260, \geq 17]]_9$.

Notice that, when providing our families of codes over $\mathbb{F}_4$, $\mathbb{F}_7$, $\mathbb{F}_8$ and $\mathbb{F}_9$, we have considered different values for the indices $j$ in order to get parameters which are either new or better than or equal to those in \cite{SLLG, cao-cui}. Finally it is worth pointing out that, when comparison is possible, our codes improve (in general, a lot) those in \cite{edel}.

\begin{rem}
{\rm
We have explained how to get $q$-ary stabilizer codes with length $nm$ by considering a class of Hermitian self-orthogonal codes of length $n$ over the field $\mathbb{F}_{q^{2m}}$, where $2 \leq n \leq q^{2m}$. Dimensions and minimum distances of the stabilizer codes depend on the $q^m$-adic expression of $n$. In certain cases, one gets better quantum codes taking Hermitian self-orthogonal codes over fields $\mathbb{F}_{q^{2m'}}$ with $m' <m$. Indeed, when the length of the quantum codes we are looking for is less than or equal to $m' q^{2m'}$ and if there exists $n' \leq q^{2m'}$ such that $nm =n'm'$, then, for distances $d \leq \min \{K_n +1, K_{n'}+1\}$ ($K_n$ and $K_{n'}$ defined as in Theorem \ref{elH} for suitable values $e=q^m$ and $e'=q^{m'}$), we obtain stabilizer codes with parameters $[[nm, nm-2m(d-1), \geq d]]_q$ if we use the extension field $\mathbb{F}_{q^{2m}}$ and  better stabilizer codes with parameters $[[n'm'=nm, nm-2m'(d-1), \geq d]]_q$ when using the extension field $\mathbb{F}_{q^{2m'}}$.
}
\end{rem}

\section*{Acknowledgments}
We thank the anonymous reviewers for their careful reading of our manuscript. We especially thank one of the reviewers for pointing out to us the existence of \cite[Lemma 76]{kkk}.

\end{document}